# Augmenting Character Designers' Creativity Using Generative Adversarial Networks.

Mohammad Lataifeh*,  Xavier Carrasco,  Ashraf Elnagar,  Naveed Ahmed

Department of Computer Science, University of Sharjah, Sharjah, 27272, United Arab Emirates

mlataifeh@sharjah.ac.ae*, xavi_2341@hotmail.com, ashraf@sharjah.ac.ae, nahmed@sharjah.ac.ae

**Abstract.** Recent advances in Generative Adversarial Networks (GANs) continue to attract the attention of researchers in different fields due to the wide range of applications devised to take advantage of their key features. Most recent GANs are focused on realism; however, generating hyper-realistic output is not a priority for some domains, as in the case of this work. The generated outcomes are used here as cognitive components to augment character designers' creativity while conceptualizing new characters for different multimedia projects. To select the best-suited GANs for such a creative context, we first present a comparison between different GAN architectures and their performance when trained from scratch on a new visual character's dataset using a single Graphics Processing Unit (GPU). We also explore alternative techniques, such as transfer learning and data augmentation, to overcome computational resource limitations, a challenge faced by many researchers in the domain. Additionally, mixed methods are used to evaluate the cognitive value of the generated visuals on character designers' agency conceptualizing new characters. The results discussed proved highly effective for this context, as demonstrated by early adaptations to the characters' design process. As an extension for this work, the presented approach will be further evaluated as a novel co-design process between humans and machines to investigate where and how the generated concepts are interacting with and influencing the design process outcome.

**Keywords:** Generative Adversarial Networks, Creative Design Process, Character Generations, Cognitive Scaffolding, Human Machine Co-creation.

## 1    Introduction

The introduction of the first GAN [1] caught the attention of many researchers due to the novel algorithmic approach it introduced. The model includes two networks competing adversely to generate new images indistinguishable from the original dataset. However, there has been no consensus on clear measures to evaluate the quality of GANs output [2], researchers often rely on Fréchet Inception Distance (FID) score [3] as an objective metric used to assess the quality of the generated images compared to ground truth. The FID score calculates the distance between two multivariate Gaussian distributions, one representing the original data, and the other for the generated output. Hence, the lower the FID score, the higher the similarities between the original and output samples. Nonetheless, the development of new and more complex architectures required new datasets as existing ones maintained low image resolutions between (32 × 32) [4] and (256 × 256) [5], [6]. Such a need motivated the creation of CelebA-HQ [7] and the Flickr-Faces-HQ (FFHQ) [8].

Consequently, more complex architectures to deal with high-resolution images also implied the demand for higher computational resources dedicated to performing the training process. Networks such as StyleGAN2-ada require 8 GPUs working in parallel for days to get the results shared in the work [9]. Despite being plausible to obtain good results with a single GPU, the consumed time and output results are far from state-of-the-art, such limitation was addressed by using a technique called transfer learning [10] and [11], which has proven to be effective when working with limited computational resources, reducing the training time and the number of features to learn.

Transfer learning is effectively used if sufficient details are made available about the previous network. Such information is shared as snap-shots or ".pkl" files or pickles that contain weights and features of networks often trained for a considerable time over large numbers of parallel GPUs, such as the case of StyleGAN2, which offers the possibility to use their pre-trained snaps for the FFHQ dataset by as a transitional starting point for the training process on a different dataset. Transfer learning will be used in this work to compare the performance and results of pre-trained pickle and newly trained models from scratch.

The remainder of the paper is organized as follows: research objectives and contributions are listed in Section 2. Related work is presented in Section 3. Section 4 describes the datasets and models used. Followed by the details of the experiments and results in Section 5, and we conclude in Section 6.

## 2  Research Objectives

Reaching lower FID scores has been one of the primary goals when developing a new GAN since it implies closer similarities, details, and fidelity of the output compared to the original dataset. However, the nature of GANs enabled further adaptation due to their ability to create something new based on a random procedure, extending their applications to different fields where novelty and creativity are sought. Indeed, innovating a new design concept for a character is much more than being novel. It must also fit a specific narrative or context. Therefore, despite the gallant strides of success in a wide range of implementations, creating a realistic result of lifelike visual scenes and characters does not work for this creative domain.

Most recent advances in other generative models such as Stable Diffusion models [12] and Semantic Image Synthesis with Spatially Adaptive Normalization [13] are deployed with open platforms such as Midjourny [2] which is raising concerns on the responsible and ethical use of these models, as thousands of designers contested the use of their creations on ArtStation [14] as training materials [15], [16] for Midjourney diffusion models. In reality, the astonishing outputs of these models are nothing but computed syntheses of human creative work. Hence, while we acknowledge the value of computational intelligence, we see this work as a proof of concept to set things in perspective on how these models can serve such a creative domain by providing cognitive aid for a human-led process.

In light of the continuous creative demand for various multimedia projects, even the most talented designers may reach a high level of exhaustion in their creative production and exhibit some limitations in their work, falling within analogical or stylistic similarities of the previously presented concepts. Designers, therefore, employ different strategies to step into fresh grounds that can inform their creative process while creating novel outputs. As such, the generated visuals in this work are proposed as a non-verbal depiction of a design brief, which is a starting point for designers to synthesize, formulate, evaluate, reflect, and create novel concepts far from being a rehash of old concepts. Furthermore, designers' perception stimulated by a visual proposition has been proven sharper compared to mental images composed from remembered representations of features, objects, structures, and semantics [17], [18].

Most of the common datasets mentioned earlier contained 30k and more images, which means a larger variety and features for the network to learn. But to provide a relevant dataset for this work, we constructed a new visual dataset to evaluate the performance of the models working with a smaller collection of images using a single GPU, this hardware limitation is not present in other works so the results obtained cannot be comparable.

Our contributions can be summarized as follows:
- Exploring the performance of different GAN architectures when trained on a context-focused dataset for characters.
- Reviewing correlations between the FID scores and human-observed perceptual quality of the generated images.

- Evaluating the performance of the models when trained on limited GPU resources compared to leveraged transfer learning.
- Paving the way toward a novel co-design process between humans and machines.

## 3    Related Work

Since its inception, the Vanilla GAN proposed in 2014 by Goodfellow [1] caught the interest of many researchers for the new algorithmic directions offered. Most of the initial improvements were related to the techniques and type of networks used in training, notably integrating the Convolutional Neural Networks [19] referred to as Deep Convolutional GAN (DCGAN); despite its distinguished output, DCGANs had three main limitations: impossibility to deal with high-resolution images, mode collapse [20] and somewhat reliance on conditioned output instead of a completely random image. Addressing the main issue of mode collapse, works like Wasserstein GAN (WGAN) [21], WGAN with Gradient Penalty (WGAN-GP) [22] or SparseGAN [23] proposed changes in the loss function and the training process, providing an effective way to avoid mode collapse, despite increasing the training time; further details on GANs and their different versions can be seen in [24], [25]

Fundamentally, the use of common datasets allowed researchers to benchmark the performance of new GANs. Early attempts were occupied with improving FID scores for the models trained on the previously noted low-resolution datasets [4]–[6]. The first successful attempt working with an image resolution of 512×512 was BIGGAN and BIGGAN-deep [26] trained on ImageNET and JFT-300M [27] datasets with significant improvements addressing mode collapse, and the ability to work with different resolutions, albeit being computationally demanding the use of at least 4 parallel GPUs. Furthermore, the introduction of the CelebA-HQ dataset in the Progressive Growing of GANs paper [28] and further implementation of this technique in StyleGAN [29] became the turning point, providing a new method to deal with high-resolution images in GANs. Additional modifications of this process were presented subsequently in StyleGAN2 [30] and StyelGAN2-ada [9].

While StyleGAN and StyleGAN2 were mainly focused on the quality of the results, there were a few datasets large enough to provide good results, the solution to this shortage came with the last model released, StyleGAN2-ada, which provides a data augmentation pipeline to avoid the discriminator over-fitting when working with little data [9], even resulting into a lower FID score, to solve the limitation caused by the number of images on which the network was trained, some techniques can be applied such as image embedding [31] to expand the latent space.

## 4       Datasets and Models

GANs create a latent space that can be freely explored after training to create a new variety of images different from the original data, allowing for flexibility and randomness in the output. To take advantage of both randomness and fidelity, we propose two consecutive pipelines shown in **Fig. 1.** and **Fig. 2**, while the first is targeting a basic silhouette output to jump-start the design process, the second one is employed to generate colored and textured alternatives for the silhouette, setting broader directions for designers on possible outcomes.

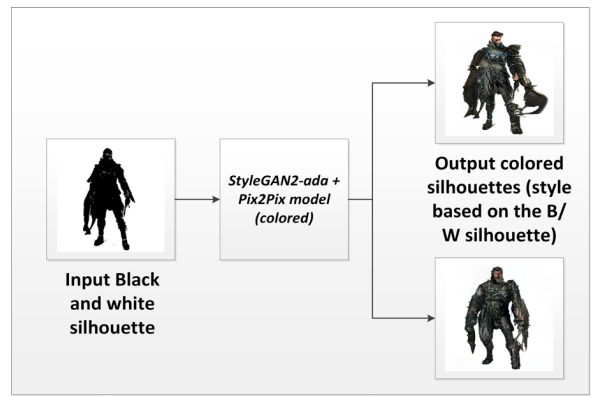

**Fig. 1.** Randomly generated character by a noise vector.

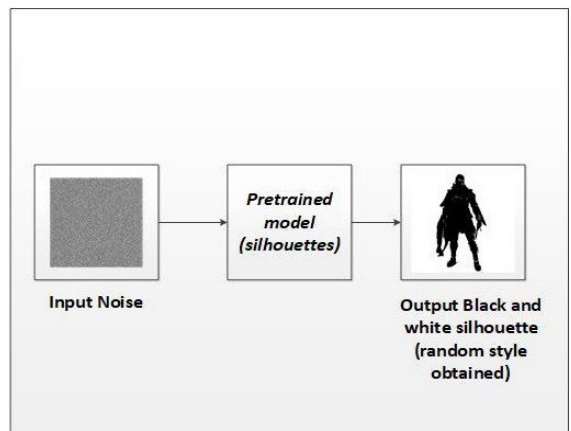

**Fig. 2.** Colored generated character based on a silhouette.

As a human-centered approach, we allow designers to evaluate the perceptual value of the generated silhouettes before proceeding to generate colored variations, this approach will be further extended in future work to a wider group of designers using a web application that allows interactions with the generative model.

### 4.1 The Datasets.

To optimize the outcomes of the selected GANs for the creative context here, a new dataset was developed and deployed for this work [32]. This was deemed necessary as the results presented in previous GANs were limited to common datasets such as CIFAR10, MNIST, LSUN, or ImageNET. While offering wide varieties, common datasets do not share enough features with our dataset in terms of context, resolution, and style.

The dataset used here included two main categories. The first one is **Character silhouettes** required to train the models for the first stage in the proposed pipeline (**Fig. 1**), allowing the generation of silhouettes from random noise. The output of this stage is carried as an input for the second stage (**Fig. 2**), during which GANs models were trained using the second category of the dataset called **Characters colored**, to generate colored/textured versions of the black/white silhouettes. Further details on the two sup-sets are provided next.

**Characters silhouettes.**
*Shape and resolution:* Squared images, resolution of 512×512. The original resolution of the images was no lower than 128×128 and they were up-sampled using a bicubic filter.

*The number of images and labeling:* The set consists of 10k images, and they are split into 3 different classes called: Man, Monster, and Woman. Some of the evaluated GANs required a unified class, hence, the images were merged into a single class.

**Characters colored.**
*Shape and resolution:* Squared images, resolution of 512×512. All images in this dataset were initially of a resolution of 512×512 or higher, they were downsampled when necessary.
*The number of images and labeling:* The set consists of 8.7k colored images and their respective silhouette version in black and white. Similar classes were used as per the first dataset.

### 4.2 The Evaluated Models.

As introduced at the beginning of this work, the proposed pipeline included two consecutive stages. The first one generates silhouettes from random noise (as shown in **Fig. 1**). Since this task is not as demanding as generating colored images, we explore the performance of different models trained on Characters' silhouettes. The models we tested for this first stage are:

- Deep Convolutional GAN (DCGAN) [19]
- Wasserstein GAN (WGAN) [21]
- WGAN with Gradient Penalty (WGAN-GP) [22]
- Large Scale GAN (BigGAN-deep) [26]
- StyleGAN2 with Adaptive Discriminator Augmentation (StyleGAN2-ada) [9]

Early GANs (DCGAN, WGAN, and WGAN-GP) share a general architecture for the generator and the discriminator, which is shown in **Table 1**. Other parameters and information related to optimizers, initializers, and the number of epochs are further detailed below. The size of the original images was also reduced to 64×64 to improve the performance of the models, since most of them are not suitable for images with higher resolutions; no other transformations were applied.

Table 1. Model for non-conditional generators and discriminators (DCGAN, WGAN, WGAN-GP).

| Operation | Kernel | Strides/ Padding | Feature Maps | Batch Norm. | Non- linearity |
|---|---|---|---|---|---|
| **Generator Input:** $G(z) - 100 \times 1 \times 1$, **z** : Input noise | | | | | |
| T. Convolution | 4× 4 | 1/0 | 64×8 | Yes | ReLU |
| T. Convolution | 4× 4 | 2/1 | 64×4 | Yes | ReLU |
| T. Convolution | 4× 4 | 2/1 | 64×2 | Yes | ReLU |
| T. Convolution | 4× 4 | 2/1 | 64 | Yes | ReLU |
| T. Convolution | 4× 4 | 2/1 | 3 | No | Tanh |
| **Discriminator Input:** $D(x) - 64 \times 64 \times 3$, **x** : Real or fake image | | | | | |
| Convolution | 4× 4 | 2/1 | 64 | No | LeakyReLU |
| Convolution | 4× 4 | 2/1 | 64×2 | Yes | LeakyReLU |
| Convolution | 4× 4 | 2/1 | 64×4 | Yes | LeakyReLU |
| Convolution | 4× 4 | 2/1 | 64×8 | Yes | LeakyReLU |
| Convolution | 4× 4 | 2/1 | 1 | No | Sigmoid |

Some important parameters modified for the models are listed below:
- DCGAN (normal and conditional): *Optimizer*: Adam ($\alpha = 0.0002, \beta_1 = 0.5, \beta_2 = 0.999$), *Loss function:* Binary Cross Entropy, Leaky ReLU slopes: 0.02, *Batch size*: 64, *Epochs*: 100, *Weight/Bias initialization type*: Uniform.

- WGAN (normal and conditional): *Optimizer:* RMSprop ($\alpha = 0.00005$), $c = 0.01$, $n_{critic}= 5$, *Leaky ReLU slopes:* 0.02,
- Batch size: 64, Epochs: 100, Weight/Bias initialization type: Uniform.
- WGAN-GP (normal and conditional): Optimizer: Adam ($\alpha = 0.0002$, $\beta_1 = 0$, $\beta_2 = 0.9$), $\lambda = 10$, $n_{critic}= 5$, Leaky ReLU slopes: 0.02, Batch size: 64, Epochs: 100, Weight/Bias initialization type: Uniform.
- BIGGAN-deep: Optimizer: Adam ($\alpha = 0.0002$, $\beta_1 = 0.5$, $\beta_2 = 0.999$), Batch size: 16, Epochs: 70, Weight/Bias initialization type: Orthogonal.

For the second stage of the proposed pipeline ( **Fig. 2**), we combined the functionalities of Pix2Pix [33] and StyleGAN2-ada [9] to color the silhouettes obtained from the previous step as well as enhance the details of the image. Both models were trained using the **Characters colored** set that was built using pairs of images of silhouettes and their respective colored versions with an original resolution of (512×512). StyleGAN2-ada was trained on the colored images only, while Pix2Pix used the colored and silhouette pairs of the set. Due to copyright limitations, we cannot share the colored images set used initially for this work. We alternatively shared a collection of 6k pairs of colored images and their respective silhouettes. The pairs were generated by using the process described here.

## 5  Experiments and Results

The main objective of this section is to present an overview of the general performance of the models introduced in the previous section. To measure the performance of these models, we calculated the FID score, followed by a human expert review for the general perceptual quality of the outcome. GANs are difficult to train due to stability issues and hardware requirements. While commonly used platforms such as Google Colab offer a wide variety of GPUs for free, these are usually randomly assigned and we could not control such allocation, so we opt for our own single GPU machines. The details of both configurations are shown below.

- CPU: Intel(R) Xeon(R) @ 2.20GHz,
- GPU: K80 or T4 or P100,
- RAM: 12 GB.

The specs of the machines used for StyleGAN-ada are:
- CPU: Intel(R) Core (TM) i7-10700K @ 3.80GHz,
- GPU: RTX 3080Ti,
- RAM: 32 GB.

CPU: Intel(R) Xeon(R) Gold 5120T CPU @ 2.20GHz, GPU: Quadro GV100 (32GB), RAM: 128 GB (only 32GB required for FID)

The software used differed according to the model. StyleGAN2-ada used Tensorflow while the others were Pytorch implementations. The requirements for DCGAN, WGAN, WGAN-GP, and BIGGAN-deep are Pytorch 1.7.1, Torchvision 0.8.2, and CUDA 11.1.

The software packages used for StyleGAN2-ada: Tensorflow 1.14, CUDA 10.0, cuDNN 7.5, Visual Studio 2015, and the VC Tools library are also required for StyleGAN2-ada.

The results for DCGAN (**Fig. 3.** ), WGAN (**Fig. 4**), and WGAN-GP (**Fig. 5**) are shown below.

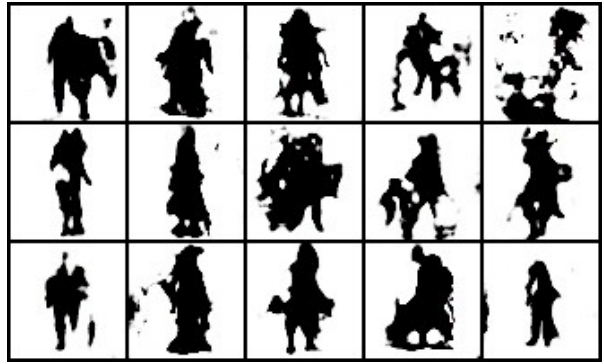

**Fig. 3.** Randomly generated character by a noise vector using DCGAN.

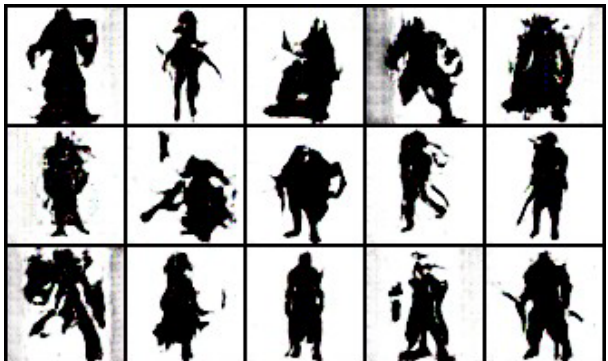

**Fig. 4.** Randomly generated character by a noise vector using WGAN.

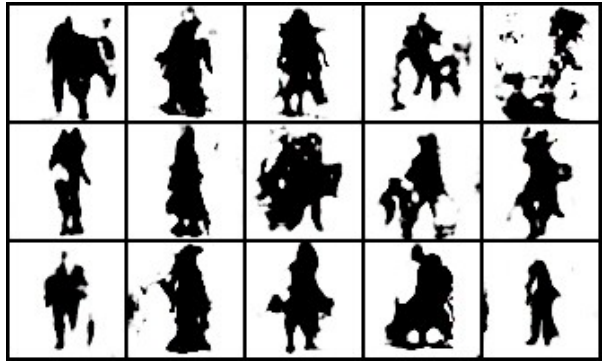

**Fig. 5.** Randomly generated character by a noise vector using WGAN-GP.

The results of the first models were obtained after a couple of hours of training, which is not comparable to the extended time required by BIGGAN-deep and StyleGAN2-ada whose generated images can be seen below in **Fig. 6** To **Fig. 9**.

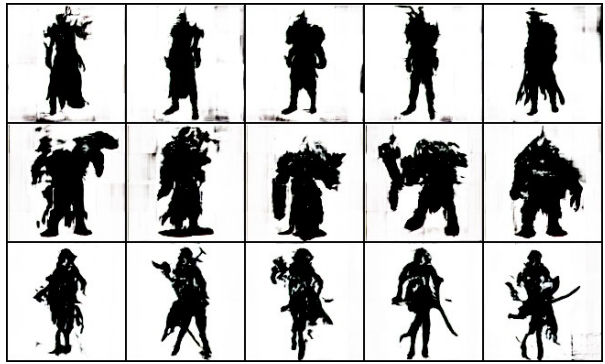

**Fig. 6.** Conditional results for BIGGAN-deep.

The generated samples for StyleGAN2-ada are split into three cases, one is for the model trained from scratch as shown in **Fig. 7.** , while **Fig. 8** and **Fig. 9** demonstrate the obtained results while using transfer learning from the tenth snap offered by NVIDIA for the FFHQ dataset. Since BIGGAN-deep was developed as a conditional architecture, we kept this feature to evaluate the results that improved significantly when compared to the previous models. Despite the shallow details, the three classes (Men, Monster, and Women) are visually distinct, with an output resolution of 128×128 pixels.

On the other hand, StyleGAN2-ada results came below expectations when trained from scratch. The classes are barely distinguishable with apparent feature similarities between the generated outcomes. Hence, the use of transfer learning was crucial to reduce processing time and improving outcome results. For the case of the exhibited outcome in **Fig. 8**, we directly used snap 10 from the pre-trained networks for FFHQ dataset. As for the results shown in **Fig. 9**, we apply different modifications to the model, such as truncation (trunc=0.75) as well as all the available augmentation techniques (–augpipe=bgcfnc), eventually, the improvements are subtle but of considerable visual value.

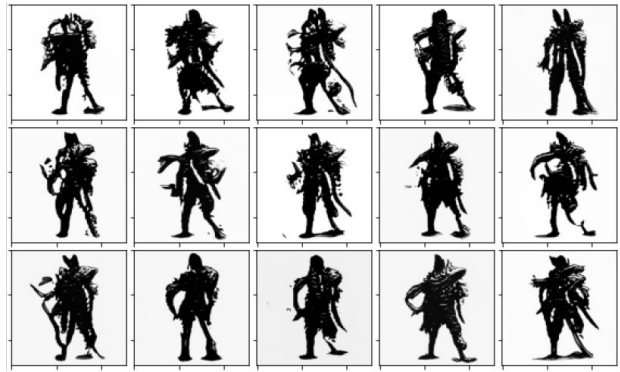

**Fig. 7.** Generated images obtained with StyleGAN2-ada trained from scratch.

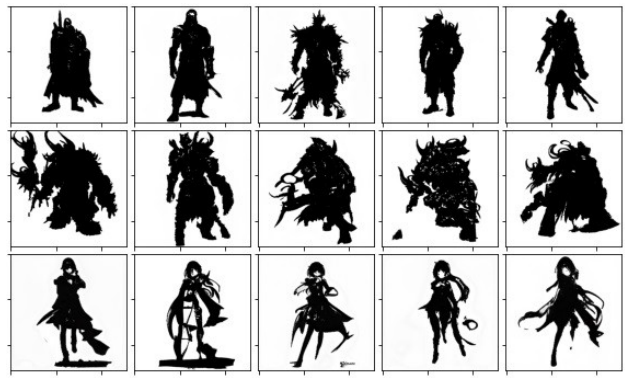

**Fig. 8.** Generated images obtained with pre-trained StyleGAN2-ad.

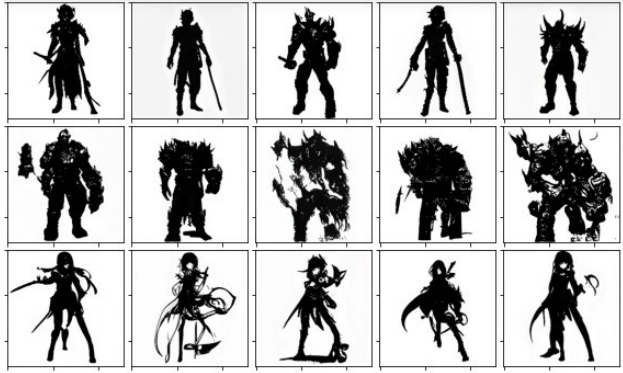

**Fig. 9.** Generated images obtained with a modified pre-trained StyleGAN2-ad.

Finally, we present the FID score for all the explored models with our dataset. The score was calculated based on 50K generated images per model. In the case of BIGGAN-deep, 16.6K images per class were generated and merged to reach 50K. The code used to calculate the FID score is a Pytorch implementation, and the obtained results are listed in **Table 2**.

Table 2. FID scores for the models.

| GAN type | FID score |
| --- | --- |
| DCGAN | 176.92 |
| WGAN | 71.25 |
| WGAN-GP | 112.59 |
| BIGGAN-deep | 47.58 |
| StyleGAN2-ada (**Fig. 7**) | 105.69 |
| StyleGAN2-ada (**Fig. 8**) | 17.60 |
| StyleGAN2-ada (**Fig. 9**) | 17.53 |

## 5.1 Designers Produced Samples

We invited character designers to review the perceptual value of the generated concepts as experts in this domain. The participating designers came with different levels of expertise, from Novice to Expert [34]. Their qualitative review of the generated concepts was critical to evaluate the value indicated by FID scores, as it also helped fine-tune the models toward the desired balance between vagueness and fidelity. The boundaries of a scope expressed in a design brief influence the perceived permissible actions, which, in this case, are depicted as a visually suggestive cognitive stimulus to entice, intrigue, and inspire.

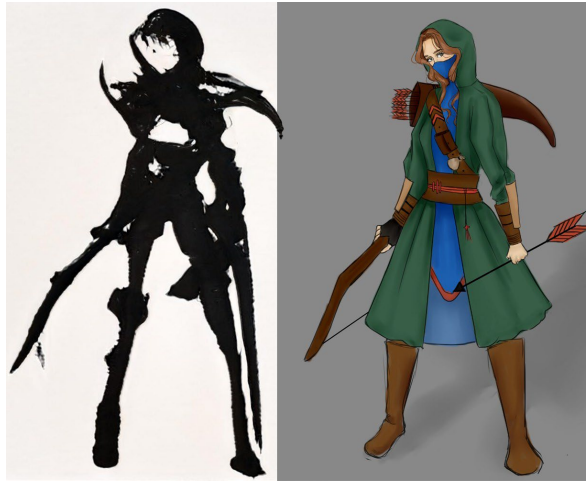

**Fig. 10.** Sample 1. Designed concept based on GANs-generated silhouettes.

A pool of selected silhouettes was made available for participants to be further developed into new concepts. The samples in **Fig. 10** and **Fig. 11** demonstrate some of the created work based on the proposed approach. Participants were asked to record their design process digitally, along with spoken narrative [35] to externalize their mental processes. The initial analysis of the collected data affirms the validity of the approach in assisting designers in creating new character concepts inspired by initial silhouettes, but novel in form, structure, and style.

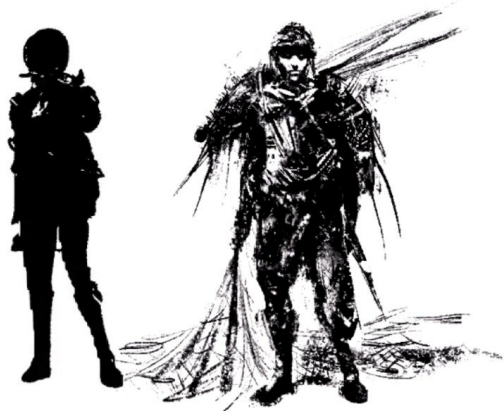

**Fig. 11.** Sample 2. Designed concept based on GANs-generated silhouettes.

Character designers described different ways of employing the generated silhouettes, from a visual design brief to metaphorical representation, inciting new directions [36]. Designers demonstrated a complex dialectical interactions [18] with the provided visuals with a verbal consensus on the uniqueness of the approach. Further analysis in future work is necessary to clarify how and where in the design process these visual cognitive elements are of the most influence, and whether such interactions will be similar for designers with different levels of expertise.

## 6      Conclusion

The advancement of GANs witnessed over the last few years has extended their value and integration to a wide range of domains and purposes. Adapting GANs output into the creative domain, this work presented the implementations of different GANs to evaluate and compare their performance when deployed with limited computational resources on a new dataset created to address the contextual need of the domain. We also explored the use of transfer learning to accelerate and ameliorate the generation process.

The early GAN models, such as DCGAN, WGAN, and WGAN-GP did not perform well considering the characteristics of the dataset and the random assignments in Google Colab GPU resources. BIGGAN-deep offered much-improved results under the same conditions. Nevertheless, the best results were obtained using StyleGAN2-ada with transfer learning, as indicated by FID scores and human expert evaluation.

While recent generative models can produce highly realistic concepts of characters, as noted for diffusion models, such output is being challenged ethically and legally as being re-syntheses of human-designed concepts. Recently, the generated concepts are denied copyright [37] claimed by their "*Engineering*" authors. Hence, we believe a co-creative design process using machine intelligence to augment human creativity sets the path to move forward.

Furthermore, we observe a positive correlation between FID scores obtained for each model with human expert evaluation, both of which are used here to ensure the developed concept will act upon their anchored value as visual cognitive elements to influence the design process. Additionally, the early work created by character designers integrating the proposed approach into their design process affirms the anticipated value of the proposed. Hence, this work not only sets a new direction for GANs applications into a unique creative domain and context, but the deployment of which is defining a novel co-design process between humans and machines. An extension of this work will further investigate the how and where of such cognitive interactions.